\documentclass[structabstract]{aa}
\usepackage{graphicx}
\usepackage{txfonts}
\usepackage{natbib}
\usepackage{hyperref}

\begin{document}
   \title{X-ray polarization fluctuations induced by cloud eclipses in active galactic nuclei}
   \titlerunning{X-ray polarimetry of cloud occultation}

   \author{F. Marin\thanks{\email{frederic.marin@asu.cas.cz}}
          \and
           M.~Dov{\v c}iak
}

   \institute{Astronomical Institute of the Academy of Sciences, Bo{\v c}ni
	      II 1401, 14131 Prague, Czech Republic}

   \date{Received 26 August 2014 / Accepted 31 October 2014}


  \abstract
  {A fraction of active galactic nuclei (AGN) show dramatic X-ray spectral changes on the day-to-week time scales
   associated with variation in the line of sight of the cold absorber. }
  {We intend to model the polarization fluctuations arising from an obscuration event, thereby offering a method of 
   determining whether flux variations are due to occultation or extreme intrinsic emission variability.}
  {Undertaking 1 -- 100~keV polarimetric simulations with the Monte Carlo code {\sc stokes}, we simulated the 
   journey of a variety of cold gas clouds in front of an extended primary source. We varied the hydrogen column
   density n$_{\rm H}$ and size of the absorber, as well as the initial polarization state of the emitting source,
   to cover a wide range of scenarios.}
  {Simulations indicate that different results are expected according to the initial polarization 
   of the extended continuum source. For unpolarized primary fluxes, large ($\sim$~50$^\circ$) variations of 
   the polarization position angle $\psi$ are expected before and after an occultation event, which is associated with very low 
   residual polarization degrees ($P~\ll$~1\%). In the case of an emitting disk with intrinsic, position-independent
   polarization, and for a given range of parameters, X-ray eclipses significantly alter the observed polarization 
   spectra, with most of the variations seen in $\psi$. Finally, non-uniformly polarized emitting regions produce very 
   distinctive polarization variations due to the successive covering and uncovering of different portions of the disk. 
   Plotted against time, variations in $P$ and $\psi$ form detectable P~Cygni type profiles that are distinctive 
   signatures of non-axisymmetric emission.}
  {We find that X-ray polarimetry is particularly adapted to probing X-ray eclipses due to Compton-thin and 
   Compton-thick  gas clouds. Polarization measurements would distinguish between intrinsic intensity fluctuations
   and external eclipsing events, constrain the geometry of the covering medium, and test the hypothesis of non-uniformly 
   emitting disks predicted by general relativity.}



\maketitle


\section{Introduction}

Since the beginning of the millennium, the number count of active galactic nuclei (AGN) known as ``changing look'' Seyfert galaxies
is increasing. More than a dozen examples present X-ray fluxes and hardness ratio light curves exhibiting large variabilities, up to a factor 3,
on year, week, or even day time scales \citep{Risaliti2002}. Such a variability pattern is attributed either to intrinsic vacillation of 
the nucleus or to transition from reflection-dominated (Compton-thick, n$_{\rm H} \ge$~10$^{24~}$cm$^{-2}$) to transmission-dominated 
(Compton-thin, n$_{\rm H} \le$~10$^{23~}$cm$^{-2}$) AGN states because of an occultation event along the observer's line of sight \citep{Risaliti2010b}. 
In this picture, discrete absorbing clouds orbit the central supermassive black hole and its surrounding emitting accretion disk, and partially 
or fully cover the central region. 

From photo-ionization, n$_{\rm H}$ and timing arguments, \citet{Risaliti2002} estimate that the absorbing material has 
to be clumpy and closer to the central engine than the so-called obscuring torus situated along the AGN equatorial plane and responsible for the 
observational dichotomy between type-1 (pole-on) and type-2 (edge-on) quasars. According to the authors, the reservoir of obscuring clouds could be 
associated with the low-ionization line broad line region (LIL BLR), where neutral or mildly ionized gas with n$_{\rm H}$ on the order
of 10$^{22 - 25~}$cm$^{-2}$ are in Keplerian motion \citep{Markowitz2014}. Such a picture is consistent with the suggestion of \citet{Netzer1993} and 
supported by the reverberation mapping observations of \citet{Suganuma2006}, implying that the LIL BLR coincides with the dust sublimation radius
of the circumnuclear torus. Observations so far point toward a distance of the eclipsing gas cloud on the order of 10$^{13 - 16~}$cm with number 
densities 10$^{8 - 11~}$cm$^{-3}$ \citep{Risaliti2009,Markowitz2014}, velocities of a few thousand kilometers per second \citep{Risaliti2009} 
and a size of the emitting X-ray source lower than 10$^{13}$cm \citep{Risaliti2009}. 

Occultation from a spherical, homogeneous cloud not only alters the X-ray light curve of obscured objects, but is also expected to affect their 
observed spectroscopic and polarimetric properties. \citet{Risaliti2011b} explored the successive shading of the receding and approaching parts 
of an AGN accretion disk, and show that an X-ray eclipse would affect the shape of the line profile and change the fluxes between the two halves of the disk 
owing to Doppler boosting. They also estimated that a 50~ks XMM-Newton or Suzaku observation could detect the asymmetries in the X-ray spectra at different 
obscuring phases of the AGN disk. When scaling down in mass, past studies of dipping\footnote{Dipping LMXRB are a subclass of X-ray binaries (XRB) known to show 
periodic dips in their X-ray light curves. Those dips are associated with obscuration by a thick gas stream located at the outer edge of the accretion disk, 
flowing from the companion star to the compact object.} low-mass X-ray binaries (LMXRB) and magnetic cataclysmic variables (mCV) have explored how occultation and 
absorption effects from the accretion stream structure affect the observed spectroscopic properties \citep{White1982}. \citet{James2002} derived the 
inclination and white dwarf mass of the mCV PQ~Gem coherently with optical polarization data. However, a polarimetric observation in only one and different 
(the optical) band is not enough for a complete interpretation, so X-ray polarimetry is needed to confirm those results. Polarization is
expected to rise since the central system, either obscured by a passing cloud at an intermediate AGN inclination or by the accretion stream in XRB, 
becomes less symmetric and offers new scattering targets.

Therefore, our aim is to present the first radiative transfer study of X-ray occultation events through the prism of X-ray polarimetry. By doing so,
we want to stress the necessity for a new X-ray polarimetric mission in order to assess some relevant questions in the field of AGN variability, (Are 
flux variations due to occultation or extreme intrinsic emission variability? Is the accretion disk emitting uniformly or, according to general 
relativity, is the emission spatially dependent?). We therefor explore the impact of a single, spherical cloud of neutral gas eclipsing an extended 
source that is representative of the irradiating accretion disk. We present in Sect.~\ref{Sec:Models} a variety of investigations, going through 
a large parameter space (energy band, hydrogen column density, size of the occulting region, intrinsic polarization level, and spatial location). 
We discuss our modeling results in Sect.~\ref{Sec:Discussion} and present our conclusions in Sect.~\ref{Sec:Conclusions}.


\section{Signatures of isolated X-ray occultation}
\label{Sec:Models}

   \begin{figure}
   \centering
   \includegraphics[trim = 12mm 92mm 15mm 1mm, clip, width=10cm]{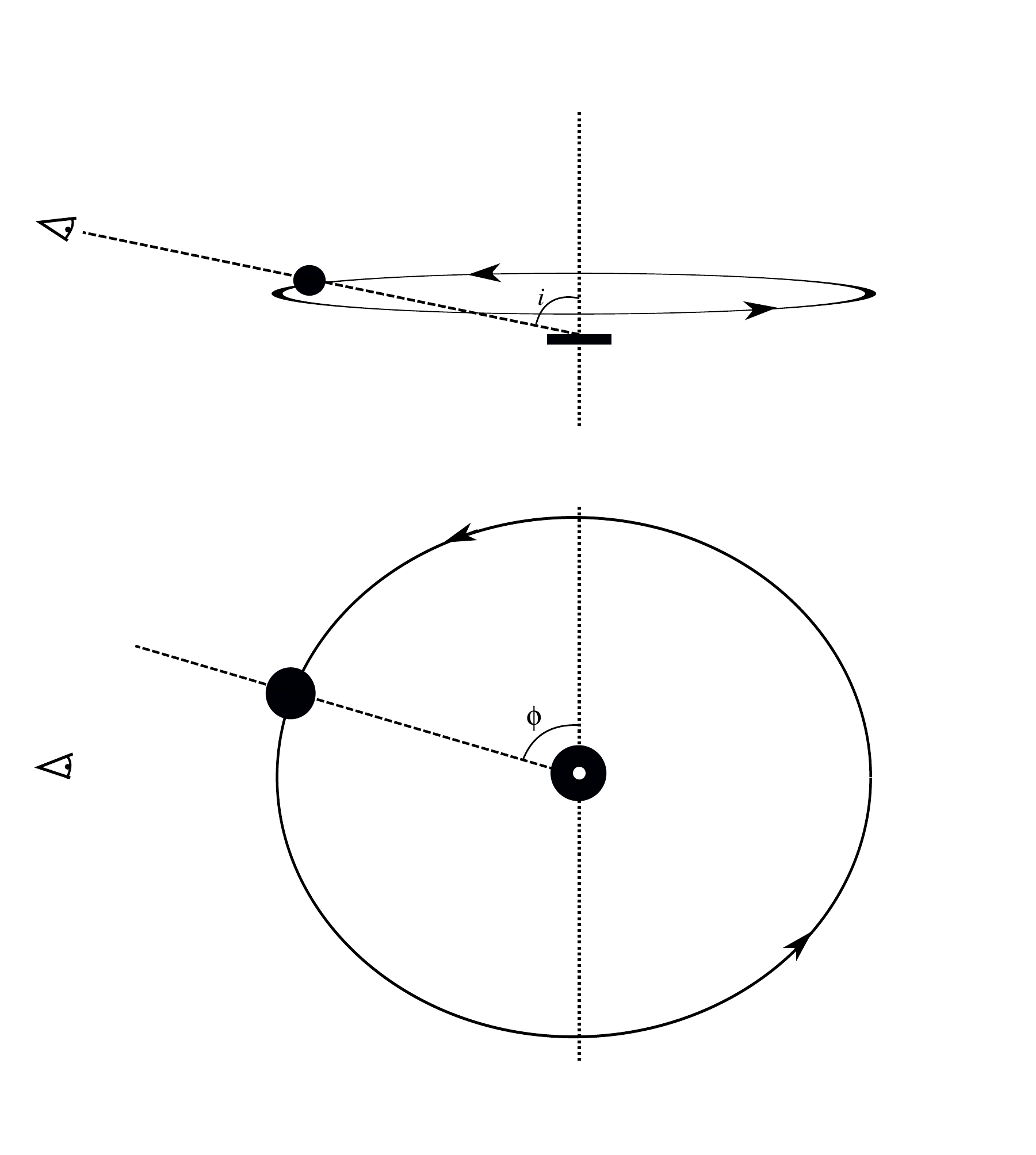}
      \caption{Representative sketch of a Keplerian cloud 
	       rotating around a central, extended source. The obscuring
	       region moves from $\phi$ = 0$^\circ$ to 180$^\circ$, 
	       with a maximum obscuration of the source at $\phi$ = 90$^\circ$.
	       The observer's viewing angle $i$ is set to 50$^\circ$.
	       The model is seen from edge-on (top figure) and pole-on (bottom) 
	       inclinations.}
     \label{Fig:Sketch}%
   \end{figure}

The physical origin of AGN X-ray light curve variations is uncertain. Several mechanisms with different physical time scales seem to superimpose: accretion-disk 
instabilities, variable accretion rates, variable obscuration of the nuclear source and/or microlensing due to stars in the host galaxy (see the reviews by 
\citealt{Ulrich1997,Peterson2001}). Since starlight is expected to show no polarization in the X-ray range (\citealt{Dolan1967}, see Sect.~\ref{Sec:Discussion}), 
most of the potential polarization comes from the disk and cloud obscuration. In this paper, we thus vary the polarization of the disk and explore the polarization 
variations due to cloud obscuration to investigate if and when the polarization signal is dominated by intrinsic emission rather than the intervening clouds.

\subsection{X-ray polarization modeling with {\sc stokes}}
\label{Sec:Models:Stokes}

Radiative transfer calculations are computed using the Monte Carlo code {\sc stokes} \citep{Goosmann2007,Marin2012}, which has already been used 
to compute the polarization emerging from a clumpy distribution of gas around the central engine of NGC~1365 \citep{Marin2013}. The code
simulates the radiative transfer between user-defined emitting and reprocessing media, computes polarization induced by multiple reprocessing 
events, and stores the final polarization properties as a function of the observer's inclination and azimuthal angle. {\sc stokes} gets 
its name from the Stokes formalism used to quantify the polarization state of each photon ($\vec S = (I, Q, U, V)^{\rm T}$). In this formalism, 
the polarization degree $P$, ranging from 0\% (depolarized) to 100\% (fully polarized), is calculated with 

   \begin{equation}
	P~=~\frac{\sqrt{Q^2+U^2+V^2}}{I} \label{ptot},
   \end{equation}

and the linear polarization position angle $\psi$, relative to the two chosen orthogonal bases of the polarization ellipse, is
   \begin{equation}
	\psi = \frac{1}{2}\tan^{-1}(\frac{U}{Q}).
   \end{equation}
   
Radiation works its way through the model region and may undergo scattering, absorption, or re-emission events. The algorithms for inelastic Compton 
scattering onto bound and free electrons, photo-ionization, and recombination effects are implemented in the X-ray regime. The emission directions, 
mean free paths between scattering events, and the reprocessing angles are computed by Monte Carlo routines based on classical intensity distributions. 
Mueller matrices are used to evaluate the change in polarization after each scattering event. Photo-absorption above the iron K-shell and the subsequent
emission of iron K$\alpha$ or K$\beta$ line photons is included and weighted against the probability of Auger effects. The scattering phase function, 
the polarization state, and the emission probability of the re-emitted photon are computed from quantum mechanics \citep{Lee1994a,Lee1994b}. {\sc stokes}
was tested against the spectropolarimetric modeling of AGN tori and found to be consistent (see discussion in \citealt{Goosmann2011}). 
For further details about the code, please refer to the complete description of the polarization properties and transformation of radiation during scattering 
events described in \citet{Goosmann2007} and references therein.

For the remainder of this paper, our fixed baseline model consists of a purely emitting, extended accretion disk with a radius $R$ of 10$^{13}$cm. The disk has a 
slab geometry with height $h \ll R$ and radiates $>$ 10$^9$ input photons according to a power-law shape $F_{\rm *}~\propto~\nu^{-\alpha}$ with a typical 
AGN spectral index $\alpha$ = 1. At a distance of 10$^{15}$cm, an atomic molecular cloud filled with neutral matter rotates in Keplerian motion around 
the source with an orbital velocity of 4000~km.s$^{-1}$. The matter of the obscuring cloud, which is assumed to be cold (i.e. $T <$ 10$^6$K), consists of neutral 
H, He, C, N, O, Ne, Na, Mg, Al, Si, S, Ar, Ca, Fe, and Ni elements with cosmic abundances. The electron optical depth of the cloud is fixed to unity (as 
may be assumed for LIL BLR clouds, \citealt{Wills1999}). The inclination of the whole system with respect to the symmetry axis of the accretion disk is 
set to 50$^\circ$, which is representative of the orientation of typical ``changing look'' AGN (e.g., NGC~1365, \citealt{Risaliti2013,Marin2013}).

Throughout the different sections of the paper, we vary several additional parameters. First, the initial polarization vector of the source will be null 
(unpolarized photons, Sect.~\ref{Sec:Models:Unpolarized}), uniformly saturated (Sect.~\ref{Sec:Models:Polarized2perc}), and then non-uniform 
(Sect.~\ref{Sec:Models:PolarizedGR}). The hydrogen column density of the gas cloud will vary from 10$^{21~}$ to and 10$^{25~}$cm$^{-2}$ to investigate 
a large panel of eclipsing schemes as reported by \citet{Markowitz2014}.

\subsection{Impact of the cloud opacities}
\label{Sec:Models:Unpolarized}

Our initial study focused on a cloud with three different hydrogen column densities (10$^{21~}$, 10$^{23~}$, and 10$^{25~}$cm$^{-2}$),
shading an extended source radiating unpolarized photons. The cloud was set to be larger than the continuum source (1.5~$\times~$R$_{\rm source}$), 
but it is slightly displaced from the observer's line of sight in order to only cover 50\% of the disk at maximum obscuration. We investigated 
the passage of the gas clump over three energy bands: 1 -- 5, 7 -- 12, and 20 -- 50~keV. We excluded the 5 to 7~keV region owing to the strong, 
broad, fluorescent iron K$\alpha$ emission\footnote{Broad Fe K-shell fluorescence lines are due to X-rays radiated from a hot, optically 
thin corona of thermally distributed electrons, which is irradiating the disk surface from above. The neutral matter of the disk absorbs and partly reprocesses 
the flux into emission lines, where the strongest is the iron K$\alpha$ line at 6.4~keV \citep{Fabian1989}. The broadening of the line is ensured by
general relativistic and Doppler effects shifting the line centroid as a function of the disk radius.} detected in a fraction of type-1 and 
intermediate AGN \citep{Nandra2007}. Such emission lines, which are unpolarized, would dilute the polarization signal. The 20 -- 50~keV band corresponds 
to the Compton hump, where Compton scattering is the most efficient way to enhance polarization. The polarization degree $P$ was integrated over the three
energy bands in order to maximize the number count of photons and thus the detectability of the polarization signal. 

   \begin{figure}
   \centering
   \includegraphics[trim = 0mm 0mm 0mm 20mm, clip, width=9.8cm]{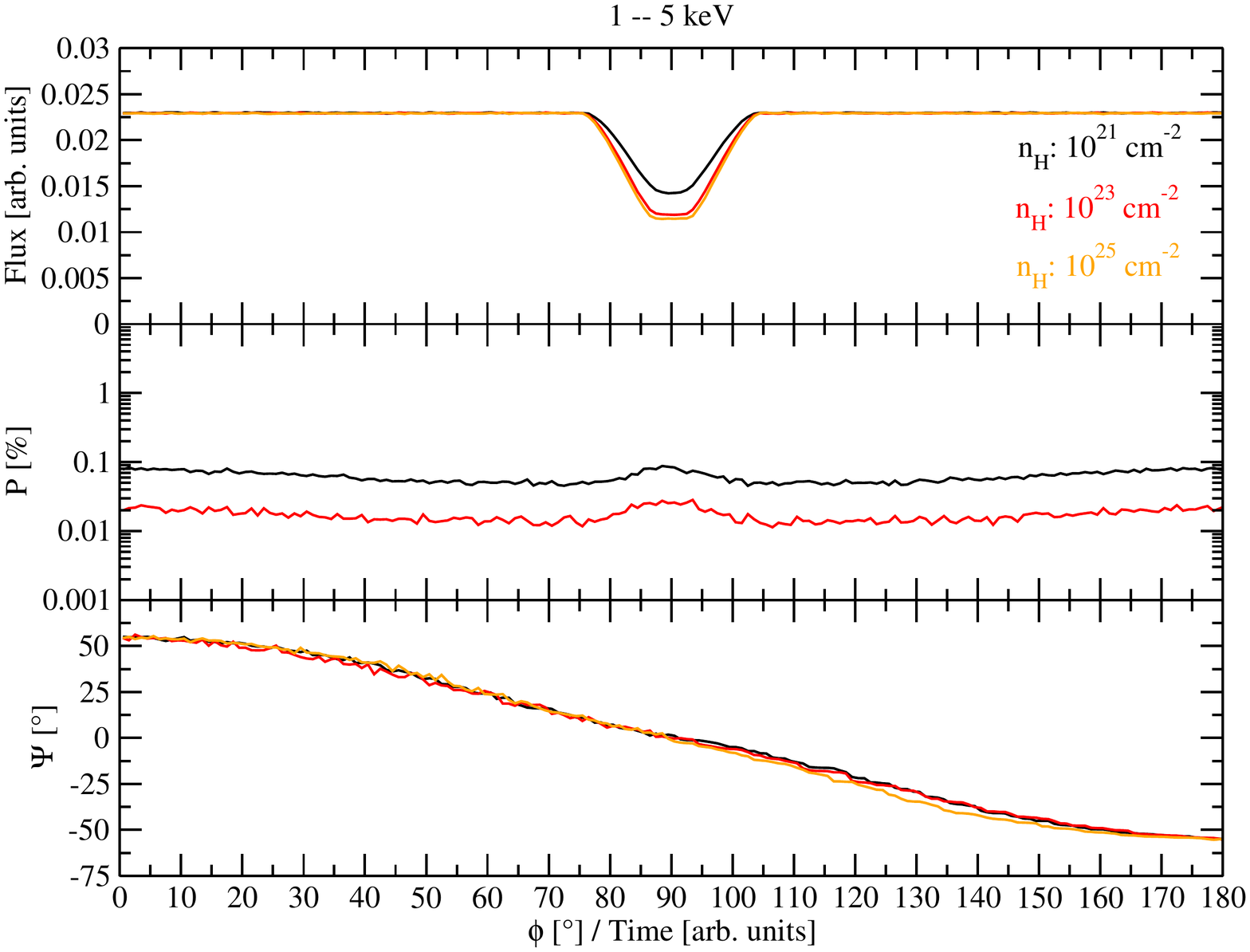}
   \includegraphics[trim = 0mm 0mm 0mm 20mm, clip, width=9.8cm]{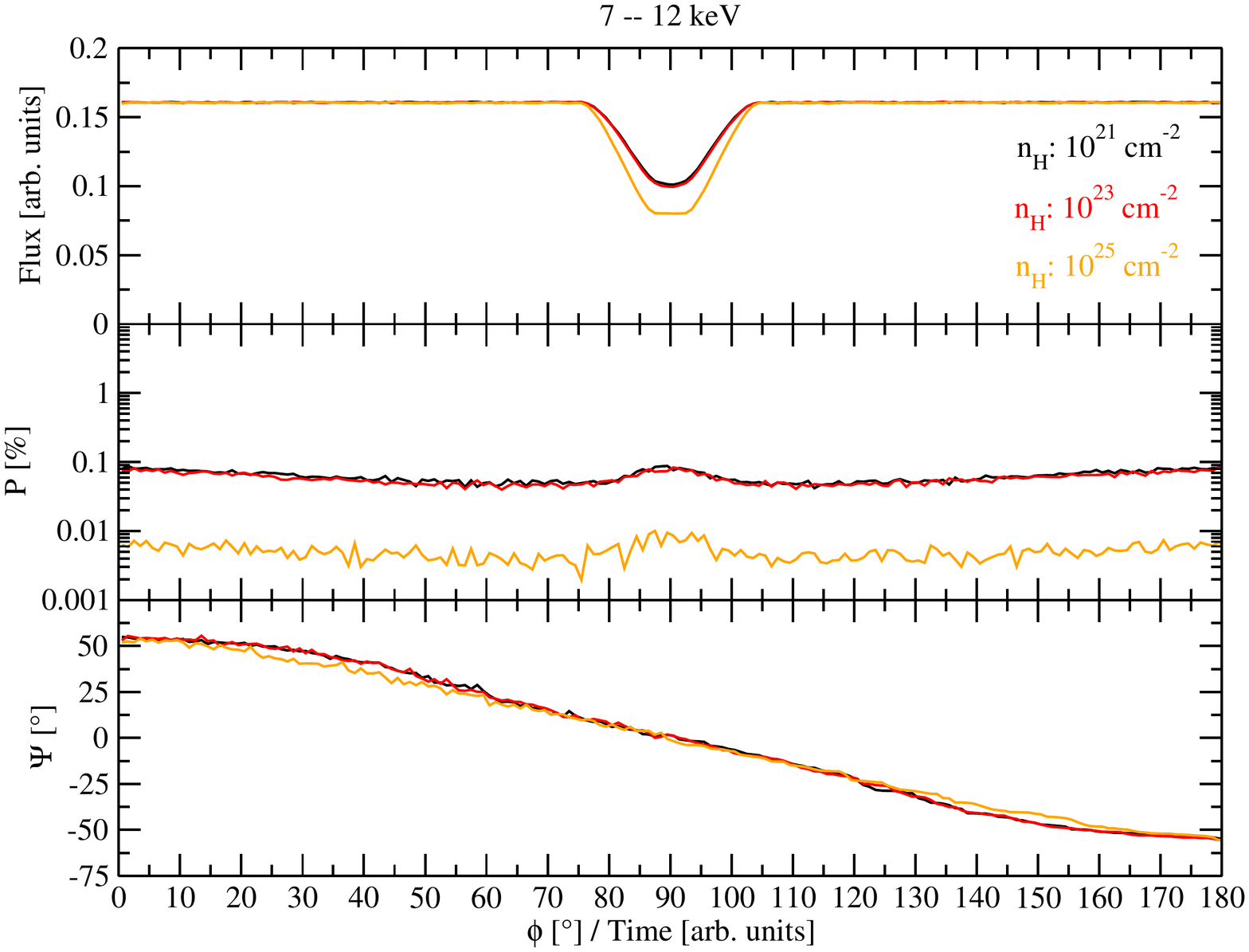}
   \includegraphics[trim = 0mm 0mm 0mm 20mm, clip, width=9.8cm]{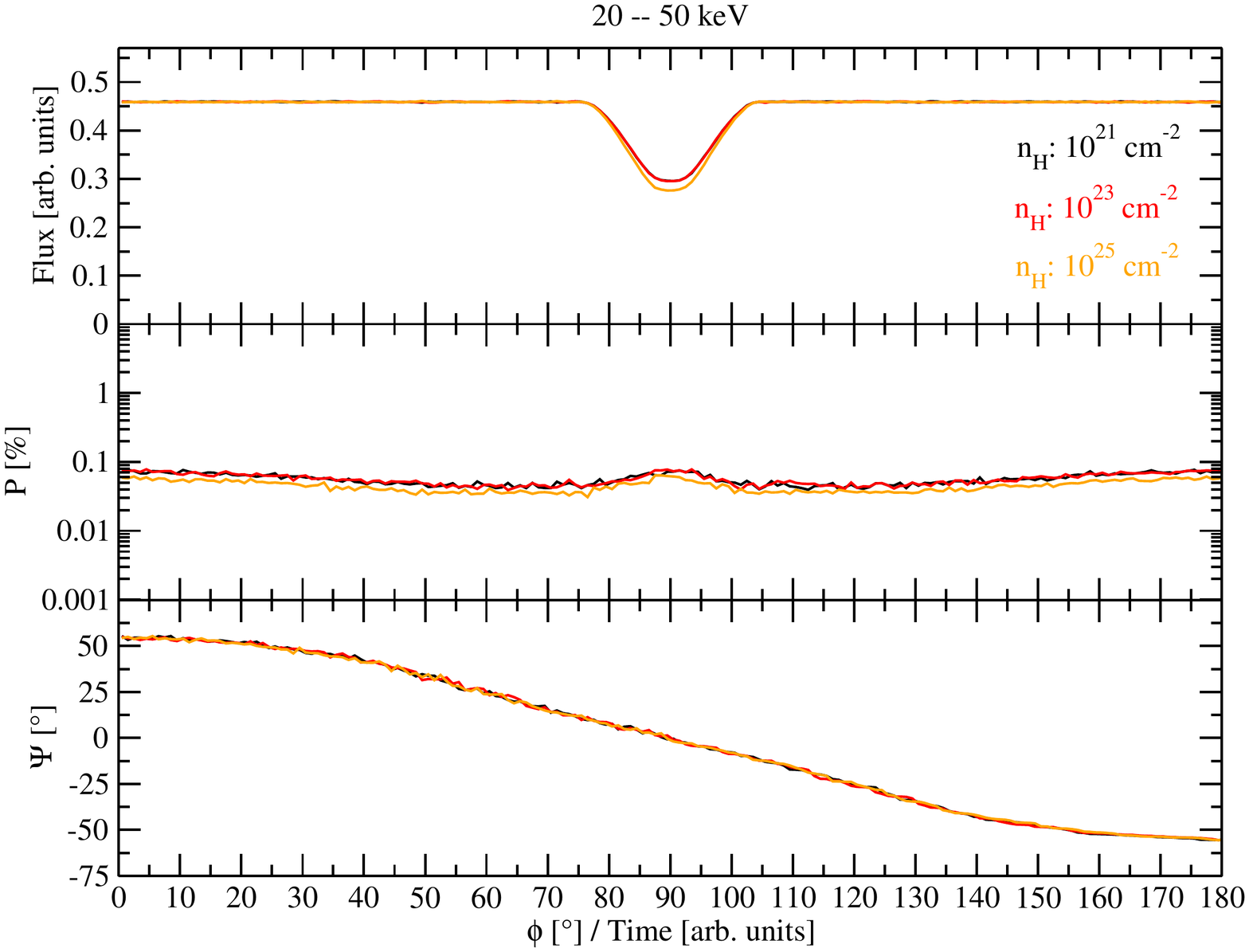}
      \caption{Time evolution of the X-ray continuum flux, polarization degree,
		and polarization position angle in the	1 -- 5~keV (top figure), 7 -- 12~keV 
		(middle), and 20 -- 50~keV band (bottom) resulting from an occultation event. 
		The eclipsing cloud covers 50\% of the extended continuum source, and 
		three different hydrogen column densities are considered for 
		the gas (black: 10$^{21~}$cm$^{-2}$; red: 10$^{23~}$cm$^{-2}$;
		and orange: 10$^{25~}$cm$^{-2}$). Since that $P_{\rm n_H = 10^{25~}cm^{-2}}$ 
		is too small, it does not appear in the top figure (see text).}
     \label{Fig:Opacities}%
   \end{figure}

We examine the time evolution of the X-ray continuum polarization resulting from an occultation event in Fig.~\ref{Fig:Opacities}.
In terms of light curve variation, occultation of the primary source is seen through a sharp decrease in the flux, which reaches
maximum attenuation when the cloud is at $\phi$ = 90$^\circ$, i.e. fully intercepting the observer's viewing angle, when 
obscuration is maximal. About half of the initial, soft X-ray flux is absorbed during occultation by a cloud with 
n$_{\rm H} \le$~10$^{25~}$cm$^{-2}$, which agrees with the XMM-Newton observation of Mrk~766 where a gas clump with 
n$_H \sim$ 10$^{23~}$cm$^{-2}$ has eclipsed the central source, decreasing the flux and the (6 -- 10~keV)/(2 -- 5~keV) hardness 
ratio light curves of Mrk~766 by 50 -- 60\% \citep{Risaliti2010b}. At n$_H \sim$ 10$^{23~}$cm$^{-2}$, absorption effects are rather 
weak and mostly apparent in the soft X-ray spectrum. At higher energies, radiation propagates through the gas cloud with even less
chance of absorption, and the three cloud models are hardly distinguishable. 

An X-ray polarization continuum appears for all n$_{\rm H}$, energies and $\phi$, even when the cloud has not yet started occulting the 
source. Scattering onto the cloud gives rise to a polarization degree that decreases smoothly with the azimuthal position 
of the cloud when departing from $\phi$ = 0$^\circ$~$\pm$~180$^\circ$. This is due to the scattering phase function of electron 
scattering that is proportional to the cosine square of the scattering angle between the source, the cloud, and the observer. 
The maximum level of $P$ emerging from this configuration is 0.1\%, which by conventional detection capabilities, is rather low: $P$ depends 
on the photon energy, the position of the cloud, and the opacity of the gas. The transit of the cloud in front of the source does not 
cancel polarization because of 1) scattering of transmitted radiation penetrating the gaseous cloud, and 2) partial shielding of 
the primary source, preventing dilution of $P$ by unpolarized radiation. The resulting polarization at the occultation point is $\sim$~0.1\%.
Polarization is extremely low for very thick gas clouds in the 1 -- 5~keV band ($\sim$~10$^{-4}$\%), since photons are heavily absorbed, 
allowing us to distinguish between Compton-thin and Compton-thick material. Polarization from dense gas clouds (n$_H$ = 10$^{25~}$cm$^{-2}$) 
rises in the 7 -- 12~keV band but still is only $\sim$~0.01\%. In the 20 -- 50~keV band, the three polarization signatures are similar 
owing to the high energies of the incoming photons, which face lower (energy-dependent) absorber scattering cross-sections. Soft X-ray polarization 
measurements are thus a better option for probing the opacity of an eclipsing gas cloud, while hard X-ray measurements would be more suitable for probing 
the intrinsic polarization properties of the source. Owing to the scattering-induced variation of $P$ and $\phi$, X-ray polarimetry predicts 
the passage of a cloud before it is detected with X-ray spectroscopy; however, such a signal is likely to be diluted by polarization by other 
LIL BLR clouds from the close environment of the central engine and thus be undetectable. 

The polarization position angle $\psi$ resulting from reprocessing by the cloud brings additional, crucial information, where $\psi$ varies with the 
azimuthal phase of the cloud from +50$^\circ$ when $\phi$ = 0$^\circ$ to -50$^\circ$ at $\phi$ = 180$^\circ$. The sequence is monotonous and 
symmetric with respect to the maximal obscuration phase, where $\psi$ = 0$^\circ$. The maximum and minimum values of the polarization position 
angle trace the non axisymmetric setup of our model and are related to the observer's viewing angle. For a cloud rotating along the equatorial
plane and  an observer situated in the same plane, $\psi$ would be equal to 90$^\circ$ (parallel polarization, measured with respect to 
the vertical axis of the system). All the gaseous clump models at all the energy ranges (even for a 10$^{25~}$cm$^{-2}$ cloud in the
1 -- 5~keV band) produce the same variation in the polarization position angle. Thus, if the initial photon flux radiated from AGN accretion 
disks is unpolarized, we expect to see variations in the measured polarization position angle before and after an occultation event.

\subsection{Occultation of a uniformly polarized source}
\label{Sec:Models:Polarized2perc}

The polarization state of X-ray radiation emerging from AGN accretion disks was investigated by \citet{Chandrasekhar1960} 
for the limiting case of $\tau~\gg$~1. The intrinsic $P$ from an optically thick, ionized, scattering slab is expected to be due to multiple 
scattering and is inclination-dependent, with a maximum of $\sim$~11\% at edge-on inclinations. The polarization position angle is fixed 
and parallel to the disk symmetry axis, independent of the inclination. In this picture, the whole disk produces the same $P$ and $\psi$.

   \begin{figure*}
   \centering
   \includegraphics[trim = 0mm 92mm 0mm 1mm, clip, width=14cm]{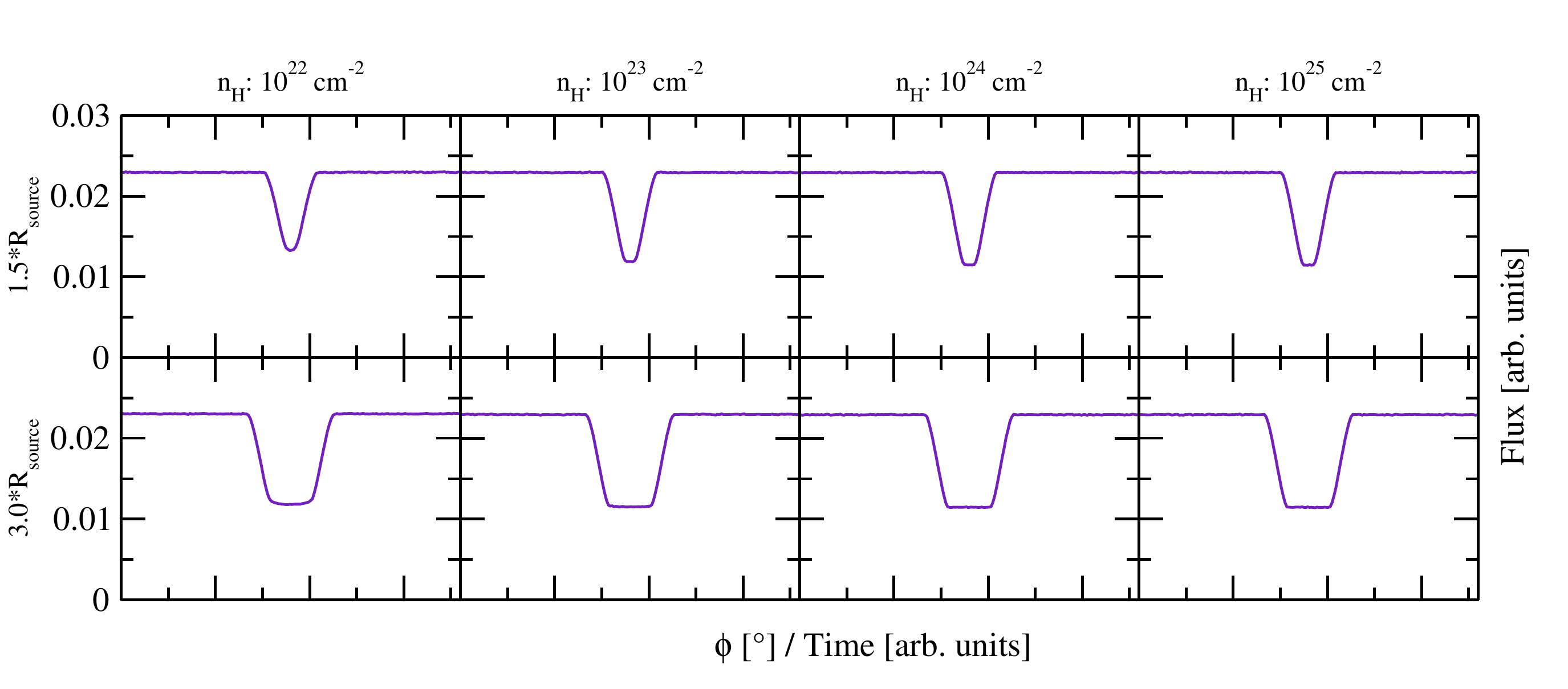}
   \includegraphics[trim = 0mm 92mm 0mm 1mm, clip, width=14cm]{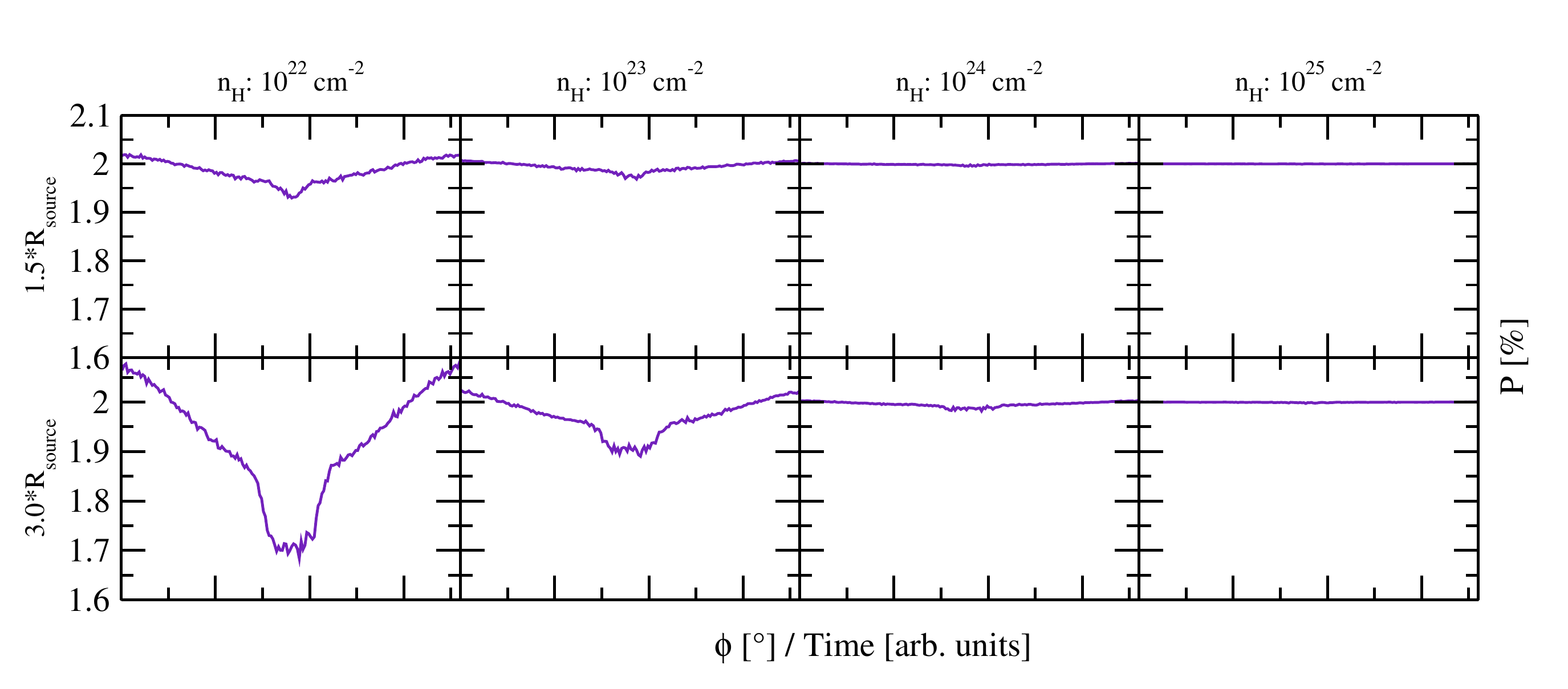}
   \includegraphics[trim = 0mm 92mm 0mm 1mm, clip, width=14cm]{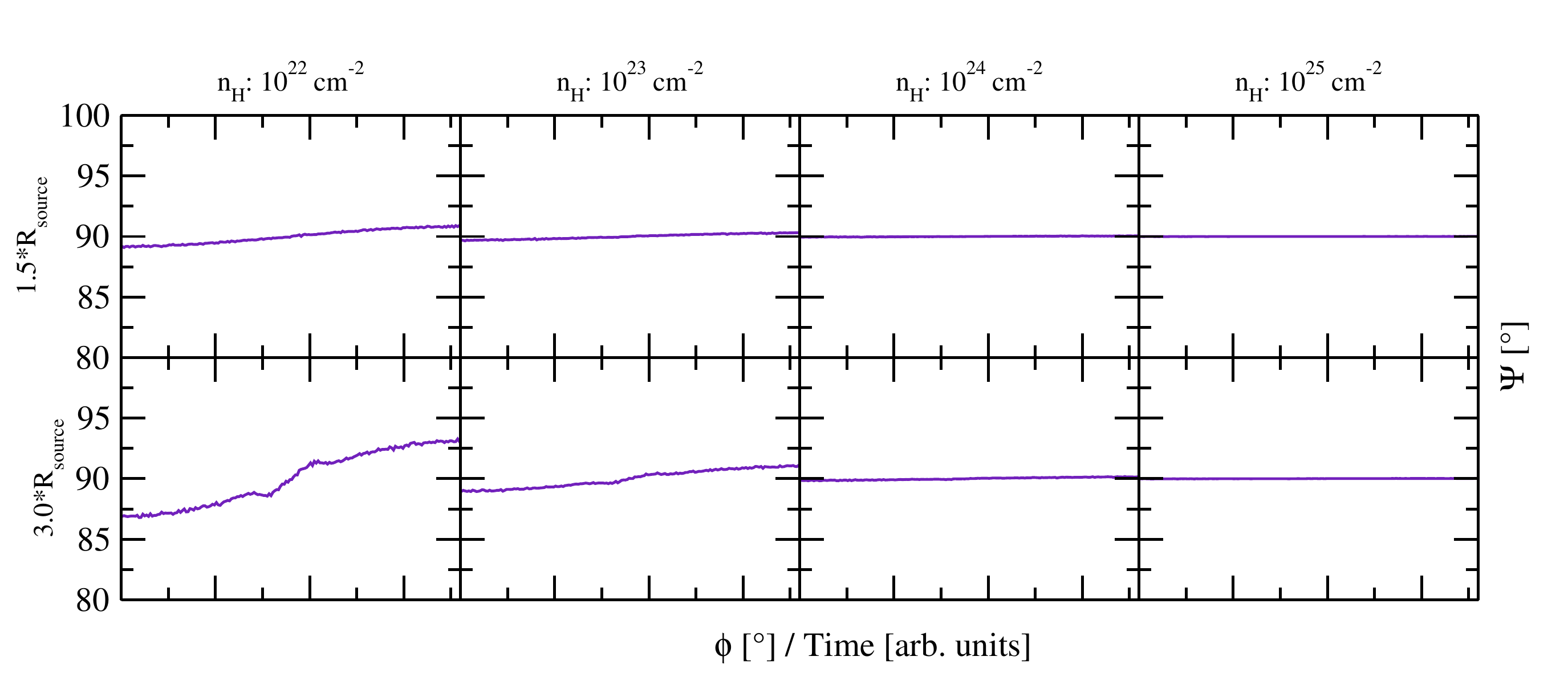}   
      \caption{Temporal evolution of the X-ray continuum flux (top figure), polarization 
		degree (middle) and polarization position angle (bottom) in the 
		1 -- 5~keV band resulting from the occultation of a primary source 
		polarized parallely to the disk symmetry axis. The cloud 
		hydrogen column density varies from 10$^{22~}$cm$^{-2}$ to 10$^{25~}$cm$^{-2}$
		(columns) and its size relative to the emitting disk ranges from 1.5 to 3 (rows);
		nevertheless, the cloud's covering factor is set to 50\%.}
     \label{Fig:GridTF}%
   \end{figure*}

Following this approach, we now undertake the exploration of a grid of uniformly emitting accretion disks (partially) covered by 
distant gas clouds. Since the inclination of the system is 50$^\circ$, the initial polarization is fixed to 2\% with a parallel polarization 
position angle, which is a polarization state representative of a reprocessing disk at an intermediate inclination \citep{Chandrasekhar1960,Angel1969,Lightman1975}.
We focus on the 1 -- 5~keV band since 1) AGN primary flux in the soft X-ray band is larger than at harder regimes, thus enhancing detection capabilities; 
2) most of the previous X-ray polarimetric mission concepts were designed to observe polarization in the soft band; and 3) we avoid 
the strong unpolarized Fe~L$\alpha_{1,2}$ ($\sim$~0.705~keV) and Fe~K$\alpha_{1,2}$ ($\sim$~6.397~keV) line complexes. We sample different hydrogen 
column densities (from 10$^{22}$ to 10$^{25~}$cm$^{-2}$) and two different cloud sizes (1.5 and 3~$\times~$R$_{\rm source}$) with a covering 
factor of 50\%. All those cases are representative of the variable absorption along the observer's line of sight in the AGN sample monitored by the Rossi 
X-ray timing Explorer (RXTE) and presented by \citet{Markowitz2014}. Their data were taken with the Proportional Counter Array (PCA) onboard the 
RXTE and limited to the 2 -- 10~keV band. They were sensitive to an absorption range of $\sim$ 10$^{22}$ -- 10$^{25~}$cm$^{-2}$, with a detection peak 
situated in the $\sim$ 4 $\times$ 10$^{22}$ -- 2.6 $\times$ 10$^{23~}$cm$^{-2}$ column density band.

The resulting grids are presented in Fig.~\ref{Fig:GridTF}. In terms of total flux, we find similar results to those in Fig.~\ref{Fig:Opacities}; i.e., 
light curve variations are governed by the hydrogen column density of the gas. Larger obscuring clouds with similar partial covering factor (50\%) tend to 
shade the continuum source longer and to create larger flux plateaus as they travel across the observer's line of sight. More significant differences 
appear in the polarization spectra: a 3~$\times~$R$_{\rm source}$ obscuring clump strongly affect the continuum $P$ at $\phi$ = 0$^\circ$~$\pm$~180$^\circ$.
If, for a 1.5~$\times~$R$_{\rm source}$ cloud with n$_{\rm H}$ = 10$^{22~}$cm$^{-2}$, $P$ is affected by scattered radiation from the cloud surface (altering $P$ 
by less than 0.08\%), then bigger clouds change the polarization degree by 0.3\% when occultation is maximum. This is due to the canceling contribution of parallel 
polarization from the disk and perpendicular polarization emerging from scattering, as seen in Fig.~\ref{Fig:Opacities}. The two polarization vectors cancel each 
other but, owing to the larger number of photons emitted from the disk, $P$ does not decrease to zero. With higher hydrogen column densities, this effect becomes 
less apparent since photons are predominantly absorbed rather than scattered. The impact of n$_{\rm H}$ and of the size of the eclipsing clump can also be seen in 
the polarization position angle spectra, because $\psi$ varies from 84$^\circ$ to 94$^\circ$ in the case of a Compton-thin absorber. This effect has better 
detection possibilities than observing a $<$~1\% change in $P$ and could help to evaluate the size of the reprocessing medium.

It is therefore found that under special circumstances (n$_{\rm H} \le$ 10$^{22~}$cm$^{-2}$ and large absorbing cloudlets), an X-ray eclipse can significantly 
alter the observed polarization spectra. In those cases, the changes in $\psi$ could help to constrain the morphology and composition of the clump. However, 
for denser clouds partially covering an extended polarized source, the signal is dominated by the intrinsic polarization of the primary radiation.

\subsection{Shading a non-uniformly polarized emitting region}
\label{Sec:Models:PolarizedGR}

According to Einstein's theory of gravity, the space time around compact objects is expected to be non-Euclidean, i.e. curved. Photons propagating 
close to the gravitational well are thus bent and move along geodesics. Computations of AGN disk irradiation by a hot corona situated above 
the disk surface, thereby producing X-ray photons via inverse Compton effect, showed that the resulting pattern is strongly anisotropic \citep{Cunningham1975}.
In addition, \citet{Connors1980} demonstrated that polarization features are also affected by general relativistic effects: the polarization vector
of light is transported in parallel along the photons' null geodesic, while the degree of polarization is a scalar invariant. Therefore, if general 
relativistic effects are shaping the radiation from accretion disks, we expect to have a non-axisymmetric, reprocessed polarization pattern.

   \begin{figure}
   \centering
   \includegraphics[width=9cm]{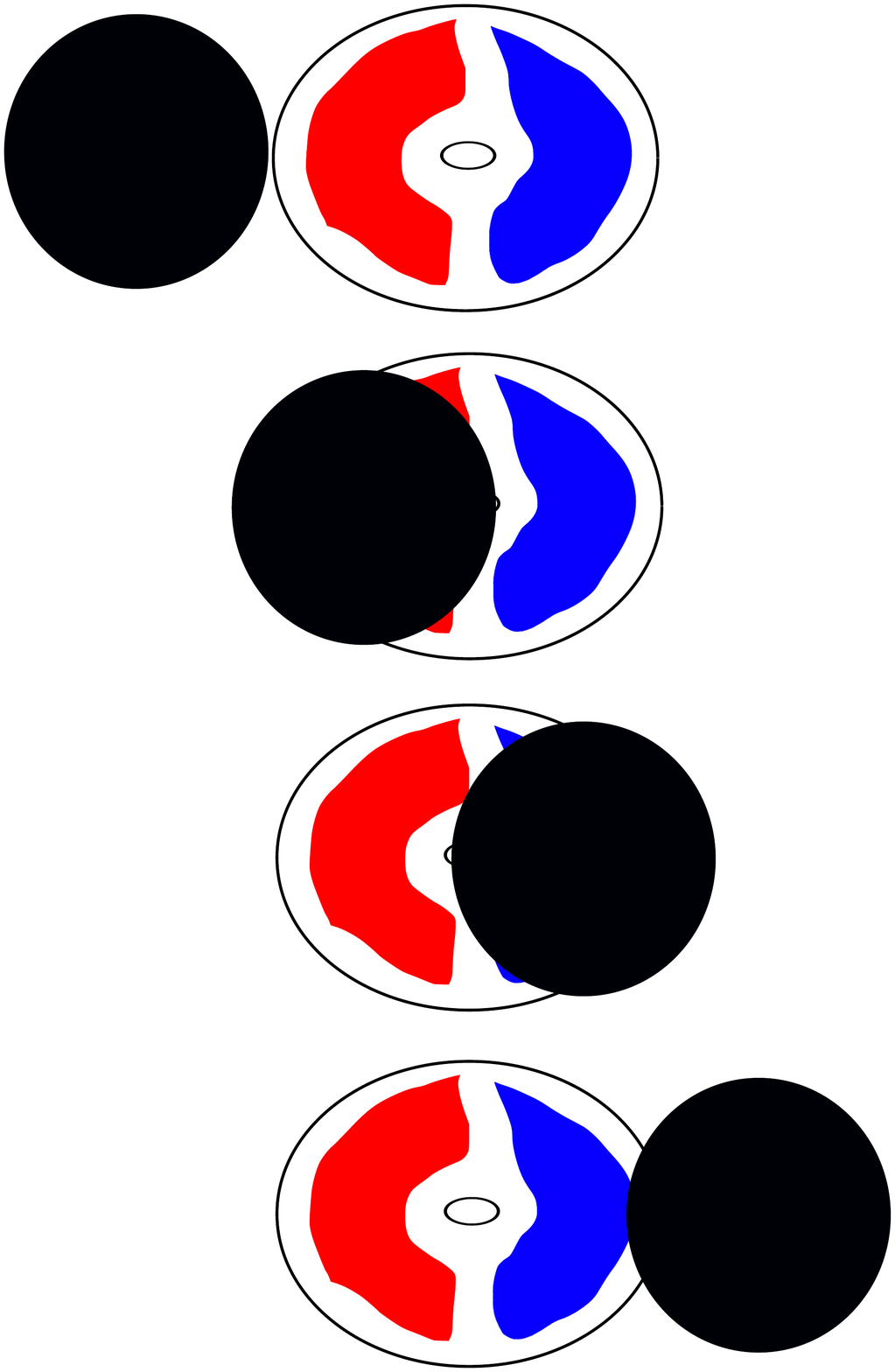}
      \caption{Illustration of an obscuring cloud shading different parts of an inclined accretion disk
	       rotating in the clockwise direction. The blue part of the accretion disk corresponds to the 
	       maximum intensity flux caused by relativistic beaming and shows 
	       a different integrated polarization state than the red part of the sketch.}
     \label{Fig:Disk_sketch}%
   \end{figure}

Similar to what has been done by \citet{Risaliti2011b} in spectroscopy, we now investigate the impact on polarization of partial disk obscuration 
by a distant gas cloud. To do so, we replace the uniformly emitting disk from our previous analysis by a toy-model disk composed of two different 
radiating parts (see Fig.~\ref{Fig:Disk_sketch}). The receding red part of the accretion disk is characterized by reprocessed radiation with an 
intrinsic polarization of 4\% associated with a 52$^\circ$ polarization position angle. The approaching blue part of the extended source emits photons 
with $P$ = 2\% and $\psi =$ 90$^\circ$. According to \citet{Cunningham1975}, the approaching and receding portions of the disk do not emit the same number 
count of photons per second. We set up three different models characterized by different luminosity ratios: 50\% (blue part) -- 50\% (red part), 
70\% -- 30\%, and 90\% -- 10\%. Here, $P$ and $\psi$ are arbitrary, but coherent representative values chosen from polarimetric simulations in strong gravity 
fields \citep{Dovciak2004,Dovciak2011,Schnittman2010}. Since the true X-ray polarization signal is likely to vary from object-to-object due to a different 
black hole mass, disk inclination, or spin, we thus investigate a generic case.

   \begin{figure}
   \centering
   \includegraphics[width=9.5cm]{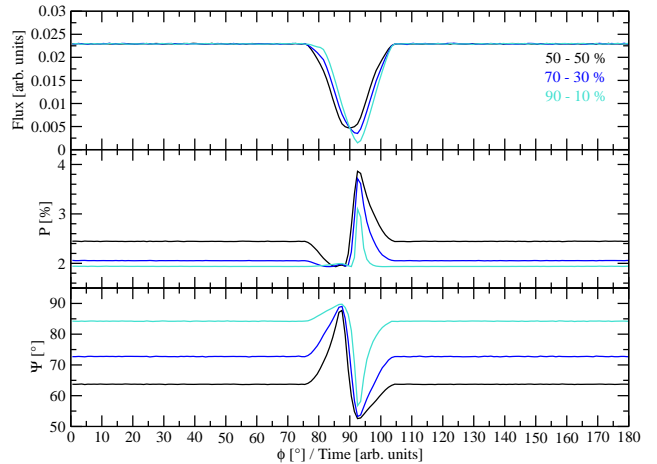}
      \caption{Time evolution of the X-ray continuum  flux, polarization degree,
		and polarization position angle in the 1 -- 5~keV band, resulting from the shading
		of different parts of a non-uniformly polarized emitting disk.
		The cloud is slightly smaller than the continuum source and has a
		hydrogen column density of 10$^{24~}$cm$^{-2}$. Three different emission
		ratios are considered with 90\% (turquoise), 70\% (blue), or 50\% (black)
		of the total luminosity being emitted by the blue part of the reprocessing disk.}
     \label{Fig:Disk_muli_polarized}%
   \end{figure}

Figure~\ref{Fig:Disk_muli_polarized} shows the variation in flux and polarization during an X-ray eclipse by a 10$^{24~}$cm$^{-2}$ gas cloud (a value similar to 
the 3.5~$\times$~10$^{23~}$cm$^{-2}$ hydrogen column density used by \citealt{Risaliti2011b}) covering about 90\% of the emitting accretion disk (ibid.).
The shape of the flux profile during an occultation event is symmetric around $\phi =$ 90$^\circ$ when the two parts of the disk emit the same number of 
photons, but departs from symmetry when the radiation pattern is affected by relativistic beaming. Covering the approaching (blue) part of the disk 
strongly affects the number of photons received by the observer and results in an asymmetric pattern, as expected by \citet{Risaliti2011b}. 
Maximum asymmetry appears for the 90\% -- 10\% model.

Polarization signatures show the most dramatic variations due to the successive covering and uncovering of the two halves of the disk. When the disk is not
obscured, the net polarization degree and polarization position angle are integrated in a way that depends on the emissivity of the approaching or receding regions.
For a disk with a 50\% -- 50\% illumination ratio, the total $P$ is equal to $\sim$~2.5\% ($\psi =$ 63$^\circ$), while for a 90\% -- 10\% model, $P \sim$~2.0\% 
($\psi =$ 84$^\circ$). Since the receding part of the disk pattern is set to emit minimum fluxes, it is natural that the integrated polarization degree and angle 
are dominated by the intrinsic polarization of the Doppler-boosted part. As soon as the red part of the accretion disk, which is responsible for the 4\% polarization 
associated with $\psi =$ 52$^\circ$, becomes obscured, the net polarization decreases smoothly to 2\% for all models. The total polarization position angle follows 
the same trend, switching to parallel polarization. Uncovering the receding red part of the disk and starting to cover its approaching blue part imprints
polarization in a way that depends more on the model. Here, $P$ rises up to 4\% and $\psi$ decreases to 52$^\circ$ for a 50\% -- 50\% scenario, but a model with higher 
approaching emissivities (e.g., 90\% -- 10\%) sees smaller $P$ variations. This is due to a large number of transmitted and scattered radiation penetrating
the gaseous cloud and altering the net polarization position angle. This effect can be seen in the polarization position angle, where the same model is found not 
to reach exactly 52$^\circ$ owing to intrinsically polarized photon transmitted through the clump and partial coverage. Once the cloud uncovers the approaching half, 
$P$ and $\psi$ return to their original, integrated values. It is noteworthy that the variation in X-ray polarization during an eclipsing event is found 
to be continuous, not sharp, forming a P~Cygni type profile\footnote{P~Cygni profiles are spectroscopic features usually detected in stellar winds and characterized 
by strong emission lines from an expanding shell with corresponding blueshifted absorption lines from material moving away from the star and toward us.} that can 
be symptomatic of general relativistic effects or, at least, of non-axisymmetric irradiation from the accretion disk.


\section{Discussion}
\label{Sec:Discussion}

In agreement with \citet{Risaliti2011b}, we have shown that X-ray eclipses affect the resulting fluxes observed by a distant observer, but we also 
provided the first time-dependent estimation of the polarimetric signatures we could expect from obscuration events. We stress that our baseline 
model provides a simple but good indicator of the polarization that a future X-ray polarimeter could observe, and we now discuss more complex cloud models and the 
possibility of adapting our work to extra-terrestrial (exoplanet) research.

\subsection{Comet-like gas structures}

Our X-ray polarimetric simulations showed that Compton-thin and Compton-thick eclipses detected in ``changing look'' AGN give different polarimetric signals 
according to the intrinsic polarization of the source and the size of the gas clump. The geometry of the obscuring cloud we used is spherical, but 
a different structure can also alter the resulting polarization spectra. From their time-resolved spectral analysis of Mrk~766, \citet{Risaliti2011} 
find hints of a more complex structure of the eclipsing cloud than a spherical clump, where strong ionization gradients point toward a comet-like geometry.
In their picture, the gas clumps are structured as follows: A dense, cold core of absorbing matter is surrounded by low-density, higher ionization layers 
of gas. The motion of the cloud is responsible for a Compton-thin cometary tail of gas extracted from the high-density cloudlet head. The resulting
morphology of the occulting medium thus departs from a spherically-symmetric model.

Since polarization is sensitive to any departure from symmetry, such a configuration is expected to increase the resulting $P$ of our simulations owing to 
additional Compton scattering within the cometary tail of the cloud. On the other hand, $\psi$ might change according to Fig.~\ref{Fig:GridTF}: the 
low-density tail of the clump may be responsible for a smooth rotation of the polarization angle after the passage of the cold head, allowing variations 
in polarization in a larger time frame. In addition, \citet{McNamara2008} shows that, for the accretion columns in mCVs, the resulting polarization 
is sensitive to the density structure of the model. In a similar way, a comet-like cloud may leave a particular imprint on polarization that would 
allow us to probe the matter and temperature stratification of those eclipsing media.

\subsection{Application to other astronomical sources}

The Keplerian orbit of a point-like object around a radiation source is reminiscent of another astrophysical domain: planetary and exo-planetary research.
By a simple analogy and dimension magnification, it is easy to consider that our gas clump can be representative of an exoplanet transiting in front of its 
parent star. We saw in Sect.~\ref{Sec:Models:Unpolarized} that when a gas clump revolves around its irradiating disk, the scattering angle changes and the 
resulting polarization varies. Scattering-induced polarization results from reprocessing of radiation onto the surface of the cloud, leading to polarization 
degrees of about 0.1\% and time-dependent $\psi$. In this case, X-ray polarimetry could be used to detect the rotation of a planet around its host. 

In the optical domain, the first direct detection of an exoplanet by polarimetric techniques was done by \citet{Berdyugina2008}, who targeted HD~189733b,
a transiting exoplanet belonging to the hot Jupiter classification, known to have a very short period (i.e., a small orbit), which makes it a valuable candidate 
for polarimetric detections. From the Stokes parameters $q$ and $u$ (normalized to the total flux), \citet{Berdyugina2008} inferred the size of the extended 
atmosphere to be on the order of 1.5 $\pm$ 0.2~R$_{\rm Jupiter}$, with a lower limit of the albedo as 0.14. In addition, the observed polarization variability 
was used to corroborate the previously estimated orbital period of the planet, inclination, eccentricity, and orientation of the orbit.

Although transparent in the optical band, the atmosphere of exoplanets is opaque at very soft X-ray energies, blocking most of the star's emission 
\citep{Poppenhaeger2013}. A model of a gas clump with n$_{\rm H} \le$ 10$^{23~}$cm$^{-2}$ and a radius smaller than the extended corona of the star could then be 
used as a first-order approximation to evaluate the resulting polarization from HD~189733b. The coronal X-ray fluxes of solar-type stars originating 
in bremsstrahlung emission and peaking in the 1 -- 2~keV band, we intend to run new Monte Carlo simulations with adapted sizes and geometries to explore such a
path (Marin \& Grosso, in prep.). Since ordinary bremsstrahlung from low-Z material is unpolarized \citep{Dolan1967}, the scattered light from HD~189733b should 
not be diluted by any intrisic polarization from its parent star and, according to Fig.~\ref{Fig:Opacities}, $P$ will be non-null. We note, though, that the 
luminosity of the parent star may be an issue for a future X-ray polarimetric detection of a hot Jupiter since the luminosity of, for example, the primary component 
of HD~189733 in the 0.25 -- 2~keV band is only on the order of L$_{\rm X} \sim$ 10$^{28~}$erg.s$^{-1}$ \citep{Poppenhaeger2013}.


\section{Conclusions}
\label{Sec:Conclusions}

This paper has shown how X-ray polarimetry, particularly in the soft, 1 -- 5~keV band, can be used to probe occulting events in
AGN. We demonstrated that the resulting polarization spectra strongly depend on the initial state of continuum 
radiation. 

In the case of an extended disk emitting unpolarized photons, a residual polarization degree appears to be due to scattering onto the cloud 
surface. Its polarization degree is low ($P~\ll$~1\%), and its polarization position angle depends on both the inclination of the system and 
on the azimuthal position of the cloud. For intermediate inclinations, $\psi$ rotates from +50$^\circ$ to -50$^\circ$, with a resulting 
polarization position angle perpendicular to the symmetry axis of the disk when the cloud is situated between the source and 
the observer.

Looking at polarized primary sources, important differences identify uniformly and non-uniformly emitting disks. When the source region 
radiates photons with a polarization state independent of the emission point, X-ray eclipses weakly impact $P$ and $\psi$ for 
Compton-thick clouds. In the case of clumps with n$_{\rm H} \le$ 10$^{22~}$cm$^{-2}$, $P$ and $\psi$ become time-dependent 
but the net variations are rather small ($\Delta P <$~1\% and $\Delta \psi <$~10$^\circ$), even when considering large obscuring clouds.
Unfortunately, such weak polarization variations are probably not detectable in the context of the present polarimetric mission 
projects.

Major differences are spotted for non-uniformly polarized emitting regions predicted by general relativity. In this case, the 
receding and approaching parts of the accretion disk are characterized by different intrinsic polarization, and the net polarization 
depends on the emissivity of the two parts. The resulting polarization variations due to the successive covering and uncovering 
of the two halves of the disk are then very distinctive: $P$ can vary by a few percent, and $\psi$ can switch from 
parallel to intermediate or perpendicular polarization according to the initial parametrization. P~Cygni type profiles are found 
to characterize the net polarization in our toy model, and similar results are expected for more complex, polarized illumination patterns.
Interestingly, several percents of variation could be in the reach of a large collecting area satellite equipped with a state-of-the-art 
polarimetric instrument if targeting a luminous object.

As a result, time-dependent polarimetry may open a new and distinctive window for probing general relativistic effects arising close to compact 
objects. In addition, polarization would help to remove the following degeneracy: Is the intensity fluctuation intrinsic to the source 
or due to an eclipsing event? If intrinsic, more chaotic variation of $P$ and/or $\psi$ are expected.

\acknowledgements This research has been partially supported by the European Union Seventh Framework Program (FP7/2013–2017)
under grant agreement no.~312789, StrongGravity. The authors would like to acknowledge the additional support from the grants COST-CZ LD12010 and 
COST Action MP1104, as well as the Academy of Sciences of the Czech Republic for its hospitality. FM is grateful to Nicolas Grosso 
and Delphine Porquet for their valuable suggestions.

\bibliographystyle{aa}
\bibliography{biblio}

\end{document}